\documentclass[a4paper,11pt]{article}
\usepackage{pos}
\usepackage{lineno}

\title{(Anti)(hyper)nuclei production in small collision systems measured with ALICE at the LHC}
\ShortTitle{(Anti)nuclei production in small systems with ALICE}

\author*[a]{Chiara Pinto for the ALICE Collaboration} 

\affiliation[a]{Technische Universitaet Muenchen,\\
James-Franck-Str. 1, Garching bei Muenchen, Germany}

\emailAdd{chiara.pinto@cern.ch}

\abstract{The production mechanism of (anti)nuclei in ultrarelativistic hadronic collisions is under intense debate in the scientific community. The description of the experimental measurements is currently based on two competing phenomenological models: the statistical hadronization model and the coalescence approach.
Light (anti)nuclei have been extensively measured in small collision systems, namely pp and p--Pb collisions, with ALICE at the LHC. Recent results on the (anti)deuteron production measured in jets and in the underlying event, and in high-multiplicity pp collisions at a center-of-mass energy $\sqrt{s}$ = 13 TeV are discussed in the context of phenomenological models, as they provide new insights on the nucleosynthesis process. }

\FullConference{%
  FAIR next generation scientists - 7th Edition Workshop (FAIRness2022)\\
  23-27 May 2022\\
  Paralia (Pieria, Greece)
}

\begin{document}
\maketitle

\section{Introduction}
\label{intro}

High-energy hadronic collisions at accelerators create a suitable environment for the production of light (anti)(hyper)nuclei. The production of light (anti)(hyper)nuclei has been extensively measured, in recent years, at various collider experiments, from the Alternating Gradient Synchrotron (AGS) \cite{AGS1, AGS2, AGS4}, to the Relativistic Heavy-Ion Collider (RHIC) \cite{RHIC1, RHIC2, RHIC3, RHIC4}, and to the Large Hadron Collider (LHC) \cite{ALICE1, ALICE2, ALICE3, ALICE4, ALICE5, ALICE6, ALICE7, ALICE8, ALICE9, jets, pp13TeVln, hyp_pPb, hyp_PbPb}. However, the production mechanism of light (anti)nuclei is not completely clear. There are two competing models used to describe the hadronization process: the thermal-statistical model and the coalescence approach. 
In the Statistical Hadronization Model (SHM) \cite{SHM}, hadrons are produced by a source in thermal and chemical equilibrium and their abundances are fixed at the chemical freeze-out. This model provides an excellent description of the measured hadron yields in central nucleus--nucleus collisions using the grand-canonical ensemble treatment \cite{SHM_2}. For small systems, such as pp and p--Pb collisions, the light nuclei production can be described using a different implementation of this model based on the canonical ensemble, where exact conservation of quantum numbers is required \cite{SHM_8}. Significant deviations in this case are observed between the experimental data and the canonical SHM predictions \cite{hyp_pPb}. The SHM is a macroscopic model, hence the mechanisms of hadron production and the propagation of loosely-bound states through the hadron gas phase are not addressed by this model. 

On the other hand, the production of light (anti)nuclei can be modelled via the coalescence of protons and neutrons that are close by in phase space at the kinetic freeze-out and match the spin, thus forming a nucleus \cite{coalescence}. 
In the coalescence model, the production probability of a nucleus with mass number $A$ is proportional to
the coalescence parameter $B_A$, which can be calculated from the overlap of the nucleus wave function and the phase space distribution of the constituents via the Wigner formalism \cite{coalescence_theory}. Experimentally, the coalescence probability $B_A$ is accessible through the ratio between the invariant momentum distribution of a nucleus with mass number $A$ and the square of the invariant momentum distribution of protons: 

\begin{equation}
B_{A} = { \biggl( \dfrac{1}{2 \pi p^{\mathrm A}_{\mathrm T}} \dfrac{ \mathrm{d}^2N_{\mathrm A}}{\mathrm{d}y\mathrm{d} p_{\mathrm T}^{\mathrm A}}  \biggr)}  \bigg/{  \biggl( \dfrac{1}{2 \pi p^{\rm p}_{\mathrm T}} \dfrac{\mathrm{d}^2N_{\mathrm p} }{\mathrm{d}y\mathrm{d}p_{\mathrm T}^{\mathrm p}} \biggr)^A},
\label{eq:BA}
\end{equation}

\noindent where the labels $A$ and p refer to the nucleus and the proton, respectively. The invariant spectra of the (anti)protons are evaluated at the transverse momentum of the (anti)nucleus, divided by the mass number $A$: \mbox{$p_{\mathrm T}^{\rm p}$ = $p_{\mathrm T}^{\mathrm A}$/$A$}. In such a model, neutrons and protons are assumed to have the same production spectra, since both belong to the same isospin doublet. 

In this paper, recent results on the production of light nuclei measured with ALICE at the LHC are discussed in the context of the production models. In particular, two recent results are highlighted in the following Sections, for their particular importance in testing the models: the measurement of the production of deuterons in high multiplicity (HM) pp collisions, and the measurement of the production of (anti)deuterons in and out of jets in pp collisions. 
In the former case, by combining the measurements of the production of (anti)nuclei and the femtoscopic measurement of the source radius in the same collision system and energy, it is possible to compare the coalescence parameter $B_2$ with parameter-free coalescence predictions. In the latter case, by comparing the (anti)deuteron coalescence probability in jets and that out of jets, it is possible to test the coalescence models, since smaller phase-space distances (and hence larger coalescence probabilities) are expected to characterize the hadrons produced in jets.


\section{Testing coalescence models in high-multiplicity pp collisions}
\begin{figure}[h]
\centering
\includegraphics[width=\textwidth]{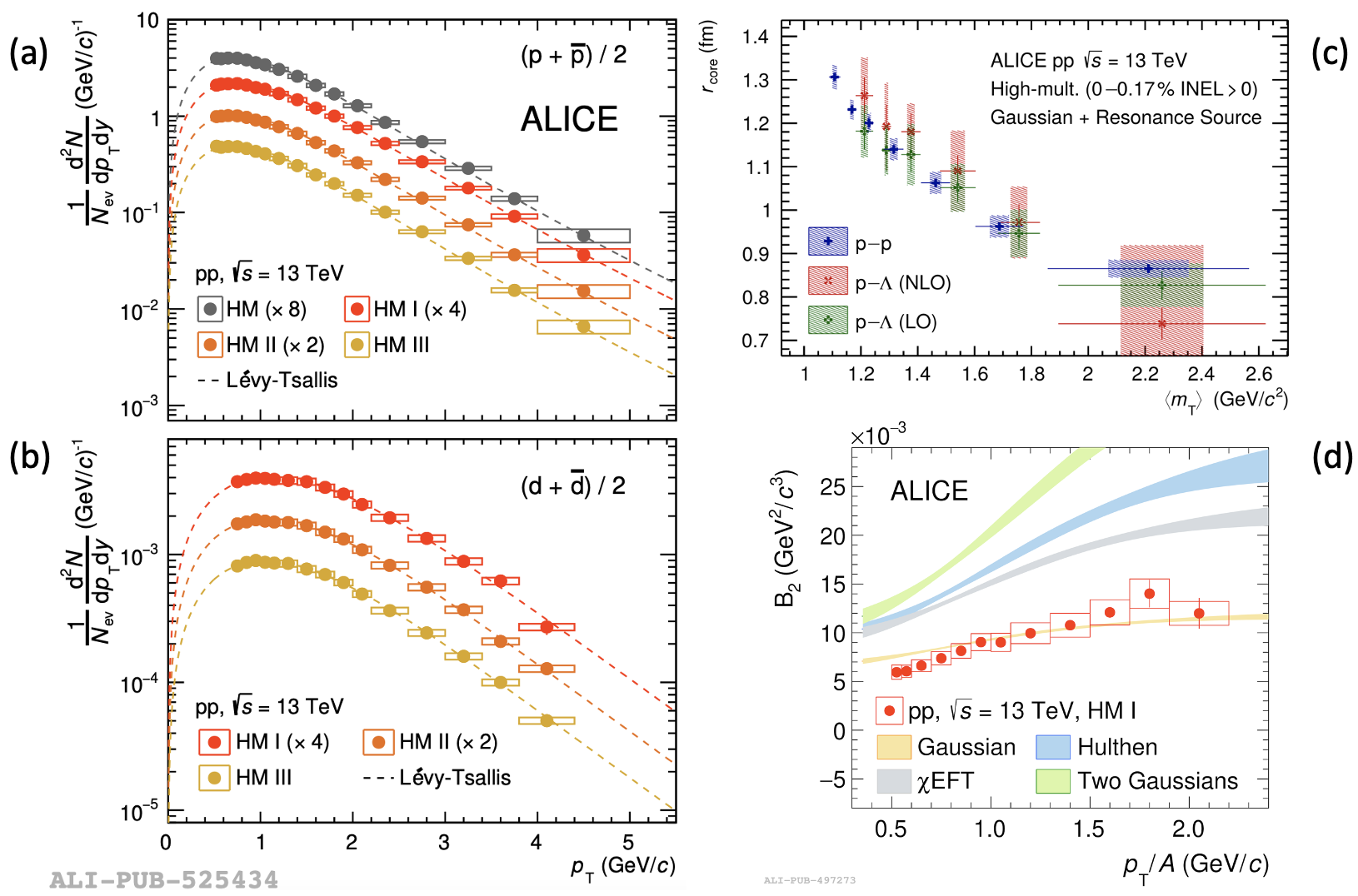}
\caption{Transverse momentum distributions of protons and deuterons measured in HM pp collisions at \mbox{$\sqrt{s}=$ 13 TeV} are shown in panels (a) and (b), respectively \cite{pp13TeVln}. The $\langle m_{\rm T} \rangle$ scaling of the source size measured in HM pp collisions at $\sqrt{s}=$ 13 TeV is shown in panel (c) \cite{HMsource}. Panel (d) shows the comparison between measurements and theoretical predictions for the coalescence parameter $B_2$ for (anti)deuterons as a function of $p_{\rm T}/A$ \cite{pp13TeVln}. Theoretical predictions are obtained using different wave functions to describe nuclei: Gaussian (yellow), Hulthen (blue), $\chi$EFT (gray) and two Gaussians (green). }
\label{fig:B2HM}       
\end{figure}

In HM pp collisions at a center-of-mass energy $\sqrt{s}$ = 13 TeV, the coalescence models can be tested by combining the measurement of the production of light (anti)nuclei \cite{pp13TeVln} with the precise measurement of the source radius with femtoscopic techniques \cite{HMsource}. In panels (a) and (b) of Fig.~\ref{fig:B2HM}, the transverse momentum spectra of protons and deuterons, respectively, needed to obtain the coalescence parameter $B_2$ with Eq. \ref{eq:BA}, are shown. Using the formalism described in Ref.~\cite{coalB2}, it is possible to compare the experimentally measured coalescence parameter $B_2$ with parameter-free coalescence predictions, where the only ingredients needed are the emission source size, which is measured for this specific dataset as shown in panel (c) of Fig.~\ref{fig:B2HM}, and the deuteron wave function~\cite{wavefunct}. As shown in panel (d) of Fig.~\ref{fig:B2HM}, several wave functions for deuterons have been tested. The Gaussian wave function provides the best description of the currently available ALICE data, despite the Hulthen one would be favoured by low-energy scattering experiments. 

\section{Testing coalescence models through production of nuclei in and out of jets}
\begin{figure}[h]
\centering
\includegraphics[width=0.7\textwidth]{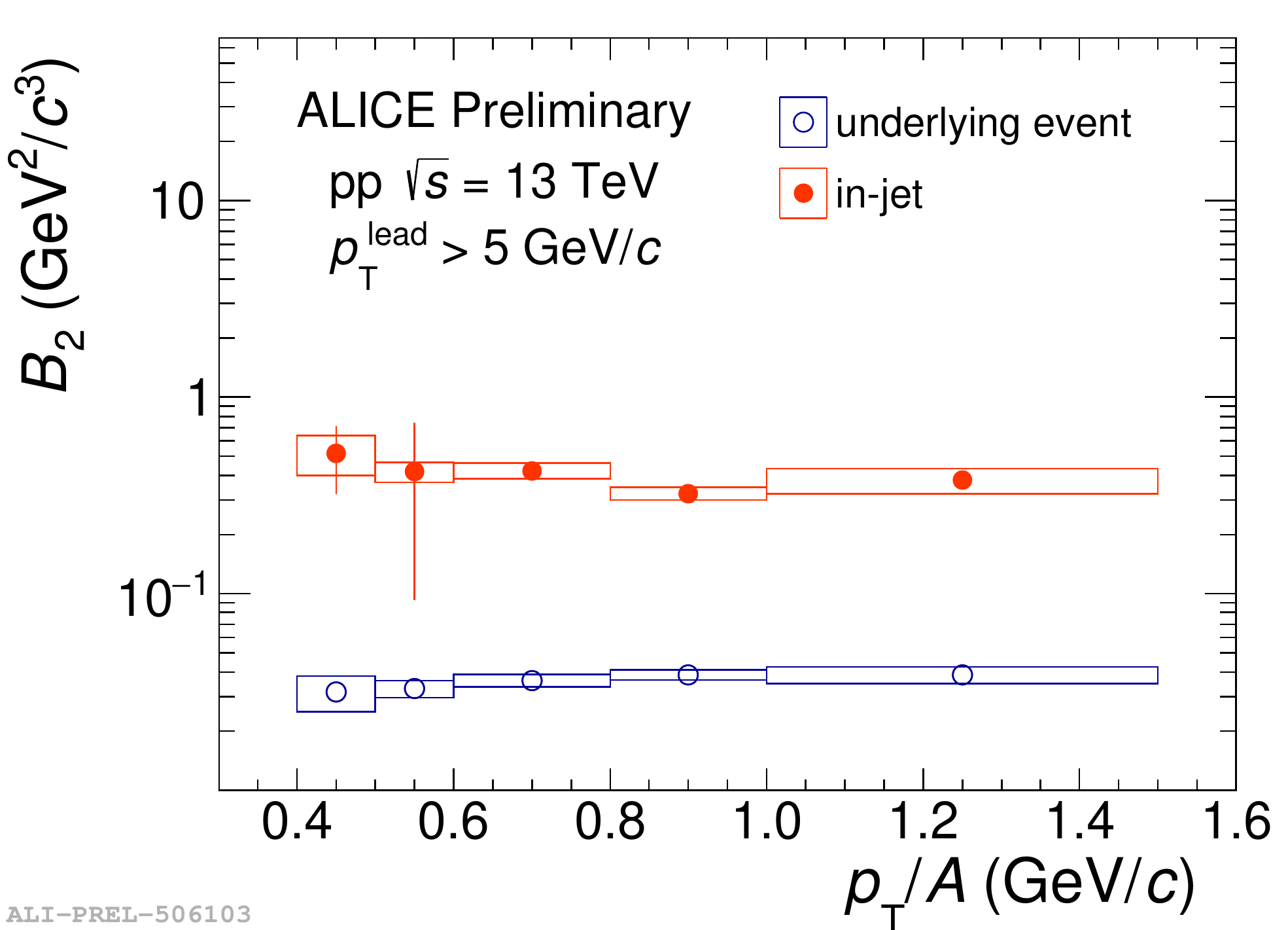}
\caption{Coalescence parameter $B_{2}$ in jets (full circles) and in the underlying event (empty circles) as a function of the transverse momentum per nucleon $p_{\rm T}/A$. }
\label{fig:B2jets}       
\end{figure}

Recent preliminary results on the deuteron production in jets and outside jets in pp collisions at $\sqrt{s}$ = 13 TeV with ALICE have shown that, despite the production of deuterons in jets is about 10$\%$ of the total production \cite{jets}, the coalescence probability in jets is one order of magnitude larger than that outside jets. 
This study has been carried out dividing the azimuthal plane in three regions, identified by the angular distance with respect to the leading particle, namely the particle with the highest transverse momentum in each event (with the condition $p_{\rm T}^{\rm{ lead}} >$ 5 GeV/$c$) and with pseudorapidity $|\eta|<$0.8. 
Three equal-size regions, 120$^\circ$ wide, are consequently defined: the one around the leading particle (Toward), the one back-to-back to it (Away), and the one transverse to both of them (Transverse). The Toward and Away regions contain contributions from the leading and recoil jets in addition to the underlying event, while the Transverse region is dominated by the underlying event \cite{UEpp}. The transverse momentum distribution of deuterons in jets is obtained by subtracting the underlying event (Transverse region) from the Toward region, resulting to be in agreement with the in-jet deuteron spectra obtained with the two-particle correlation method \cite{jets}, while the underlying event transverse momentum distribution corresponds to the Transverse region one. 
The coalescence parameters $B_2$ in jets and in the underlying event are obtained as a function of $p_{\rm T}/A$ using Eq. \ref{eq:BA} and shown in Fig. \ref{fig:B2jets}. The observed enhancement of the in-jet $B_2$ is in agreement with the coalescence picture, and it can be interpreted as due to the reduced distance in phase-space between nucleons in jets compared to hadrons outside of jets. These results will be compared with predictions from simple coalescence and from a reaction-based deuteron production model. In the former case, the transverse momentum distributions of nucleons are generated using PYTHIA 8 with Monash 2013 tune, while in the latter case (anti)deuterons are generated by ordinary nuclear reactions between the nucleons produced in the collision with parametrized energy-dependent cross sections tuned on available experimental data. 

\begin{thebibliography}{}

\bibitem{AGS1}
E878 Collaboration, Phys. Rev. C 58 (1998) 1155-1164.
\bibitem{AGS2}
E802 Collaboration, Phys. Rev. C 60 (1999) 064901.
\bibitem{AGS4}
E864 Collaboration, Phys. Rev. Lett. 85 (2000) 2685–2688.
\bibitem{RHIC1}
STAR Collaboration, Phys. Rev. Lett. 87 (2001) 2623011–2623016.
\bibitem{RHIC2}
PHENIX Collaboration, Phys. Rev. Lett. 94 (2005) 122302.
\bibitem{RHIC3} 
BRAHMS Collaboration, Phys. Rev. C 83 (2011) 044906.
\bibitem{RHIC4}
STAR Collaboration, Nature 473 (2011) 353 [Erratum: Nature 475 (2011) 412].
\bibitem{ALICE1}
ALICE Collaboration, Nature Phys. 11 no. 10, (2015) 811–814.
\bibitem{ALICE2}
ALICE Collaboration, Phys. Rev. C 93 no. 2, (2016) 024917.
\bibitem{ALICE3}
ALICE Collaboration, Eur. Phys. J. C77 no. 10, (2017) 658.
\bibitem{ALICE4}
ALICE Collaboration, Phys. Rev. C 97 no. 2, (2018) 024615.
\bibitem{ALICE5}
ALICE Collaboration, Nucl. Phys. A 971 (2018) 1–20.
\bibitem{ALICE6}
ALICE Collaboration, Phys. Lett. B 794 (2019) 50–63.
\bibitem{ALICE7}
ALICE Collaboration, Phys. Lett. B 800 (2020) 135043.
\bibitem{ALICE8}
ALICE Collaboration, Phys. Rev. C101 no. 4, (2020) 044906.
\bibitem{ALICE9}
ALICE Collaboration, Eur. Phys. J. C80 no. 9, (2020) 889.
\bibitem{jets}
ALICE Collaboration, Phys. Lett. B 819 (2021) 136440.
\bibitem{pp13TeVln}
ALICE Collaboration, JHEP 01 (2022) 106.
\bibitem{hyp_pPb}
ALICE Collaboration, Phys. Rev. Lett. 128 25 (2022) 252003.
\bibitem{hyp_PbPb}
ALICE Collaboration,  Phys. Lett. B 754 (2016) 360-372. 


\bibitem{SHM}
A.Andronic, P.Braun-Munzinger, J.Stachel, H.Stöcker, Phys. Lett. B \mbox{697 (2011) 203-207}. 
\bibitem{SHM_2}
A.Andronic, P.Braun-Munzinger, K.Redlich, J.Stachel, Nature 561 (2018) 321–330.
\bibitem{SHM_8}
V. Vovchenko, B. Doenigus, H. Stoecker, Phys. Lett. B 785 (2018) 324.
\bibitem{coalescence}
Butler, S. T. and Pearson, C. A., Phys. Rev. 129 (1963) 836. 
\bibitem{coalescence_theory}
R. Scheibl and U. W. Heinz, Phys. Rev. C59 (1999) 1585-1602.
\bibitem{HMsource}
ALICE Collaboration, Phys. Lett. B 811 (2020) 135849.
\bibitem{coalB2}
K. Blum, M. Takimoto, Phys. Rev. C 99 4 (2019) 044913.
\bibitem{wavefunct}
F. Bellini et al., Phys. Rev. C 103 (2021) 014907.
\bibitem{UEpp}
ALICE Collaboration, JHEP 04 (2020) 192.
\end{thebibliography}
\end{document}